# Surface tension-driven convection patterns in two liquid layers

Anne Juel,[1] John M. Burgess, W.D. McCormick,
J.B. Swift, and Harry L. Swinney[2]

*Center for Nonlinear Dynamics and Department of Physics,*
*The University of Texas at Austin,*
*Austin, Texas, 78712 USA*

**Abstract**

Two superposed liquid layers display a variety of convective phenomena that are inaccessible in the traditional system where the upper layer is a gas. We consider several pairs of immiscible liquids. Once the liquids have been selected, the applied temperature difference and the depths of the layers are the only independent control parameters. Using a perfluorinated hydrocarbon and silicone oil system, we have made the first experimental observation of convection with the top plate hotter than the lower plate. Since the system is stably stratified, this convective flow is solely due to thermocapillary forces. We also have found oscillatory convection at onset in an acetonitrile and n-hexane system heated from below.

## 1 Introduction

When a liquid layer with a free surface is heated from below, both surface tension gradients and buoyancy may drive convective motion (1), but thermocapillary forces are typically dominant when the depth of the liquid layer is small. This phenomenon is known as Bénard-Marangoni convection and its typical signature is an hexagonal array of convection cells that develops above a critical temperature difference. The bifurcation from conduction to convection is subcritical (2; 3) and secondary transitions are reported to both steady (4) and unsteady (5) square planforms. In the limit of very thin liquid films, a long wavelength instability is favored which is deformational in nature (6).

---

[1] Electronic address: aj@chaos.ph.utexas.edu
[2] Electronic address: swinney@chaos.ph.utexas.edu

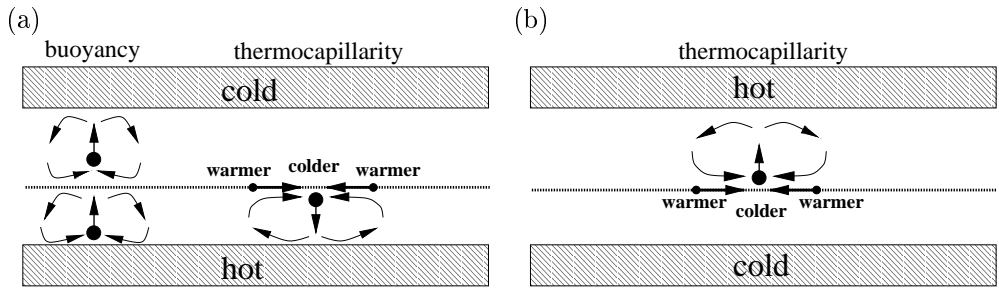

Fig. 1. Schematic diagram of the destabilizing mechanisms in two liquid layer Bénard-Marangoni convection. (a) When the system is heated from below, thermocapillary forces are destabilizing in the bottom layer and stabilizing in the upper layer, while buoyancy forces are destabilizing in both layers. (b) When the system is heated from above, thermocapillary forces are destabilizing in the upper layer and stabilizing in the bottom layer, while buoyancy forces are stabilizing in both layers.

Analogous systems with an interface between two liquids have received considerably less attention. In the traditional system, the gas layer above the liquid may be treated as passive [3], whereas a full two-layer description is required for liquid-liquid systems. The variety of convective mechanisms is greatly enhanced in this more general configuration due to the interaction between the flows in each of the superposed liquid layers (7). For instance, the flow can be either viscously coupled, so that superposed convection rolls are counter-rotating, or thermally coupled. In this case, superposed rolls are co-rotating, so that a local recirculation is required to satisfy the interfacial boundary condition. The increased complexity of the two liquid-layer system leads to instabilities inaccessible in the liquid-gas system.

Convection may arise in a liquid-liquid system when the temperature difference is applied by heating from above, due solely to thermocapillary forces since the liquid layers are stably stratified. A comparison between the instability mechanisms for Bénard-Marangoni convection heating from below and above is shown in Fig. 1. In Fig. 1(a), the bottom plate is hotter than the top plate, so when a fluid particle close to the interface in the lower layer is displaced downwards due to a natural perturbation, warmer fluid flows inwards to replace it by continuity, thus creating two hotter spots on the interface on either side of a colder region. Since the temperature derivative of the interfacial tension is negative, these local gradients can drive convective motion if viscous and thermal dissipation are overcome. Thus, thermocapillary forces are destabilizing in the lower layer and so are buoyancy forces in both layers since the system is unstably stratified. Also, by the same mechanism, thermocapillary forces in the upper layer are stabilizing. When the top plate is hotter than the bottom plate as shown in Fig. 1(b), the situation is reversed. Thermocapillarity is destabilizing in the upper layer and stabilizing in the lower layer. Buoyancy is now stabilizing in both layers, since the system is stably stratified. Hence, linear stability calculations are required to predict the specific parameter range where the instability occurs.

---

[3] For 10 cS silicone oil and air, the density of the gas is 3 orders of magnitude smaller than that of the liquid, and the thermal diffusivity of air is 200 times larger than that of the silicone oil, while their kinematic viscosities are of the same order of magnitude.

Zeren and Reynolds (9) theoretically predicted convection heating from above for a water and benzene system, but were not able to observe it experimentally. They attributed its absence to the existence of impurities at the interface.

New convection patterns have been found in a liquid-liquid configuration heated from below, as reported by Tokaruk *et al.* (10), who observed a square planform at onset and then a secondary transition to rolls in a perfluorinated hydrocarbon and water system. In their experiment, the destabilizing effects of buoyancy and thermocapillarity were of the same order.

In the two liquid layer system, heated either from above or below, the convective motion at onset can be oscillatory, in contrast to single-layer Rayleigh-Bénard systems where oscillatory convection is only found as a secondary instability (11; 12). Although Takashima (13) predicts the onset of convection in a liquid layer with a free surface heated from above, due to a deformational instability, the required critical temperature differences are too large to be realized in practice.

In two superposed liquid layers, several mechanisms can lead to oscillatory convection. The situation where thermocapillarity is negligible has been investigated both theoretically and experimentally (14; 15; 16); here thermal and viscous couplings between the convection rolls in each layer can compete to generate time-dependent behavior. When thermocapillarity is dominant, the competition between stabilizing and destabilizing forces in each layer may also lead to oscillatory convection at onset (1). However, this has not been observed experimentally.

The linear stability of the two liquid layers has been investigated theoretically by several authors (9; 17; 18), and Simanovskii and Nepomnyashchy (18) include the study of stability with respect to time-periodic perturbations. A weakly nonlinear stability analysis was derived by Golovin *et al.* (19) for a liquid-gas system where the upper layer had negligible viscosity, and Engel and Swift (20) recently extended this work to the more general problem of steady flow in two liquid layers. The large number (10 if the interface is indeformable) of control parameters governing the dynamics prohibits an exhaustive theoretical investigation of linear solution classes. However, once the liquids have been selected, the depths of the layers and the applied temperature difference are the only independent control parameters.

In this paper we examine convection patterns in liquid combinations where buoyancy and thermocapillary forces both contribute to the instability. In Section 2 we describe the linear stability analysis of the full two-layer problem, which guides our experimental search, and we report specific predictions for the phenomena new to the liquid-liquid system. The experimental apparatus is described in Section 3 and experimental results are discussed in Section 4. In particular, we report the first experimental observation of Marangoni instability with heating from above and the first experimental evidence of the onset of standing waves as a primary instability in a system where both thermocapillarity and buoyancy are important. Additional phenomena associated with inter-diffusion of the liquids, which is unavoidable even for nominally immiscible liquids, impair the study of the convective phenomena. These effects are discussed in Section 4.4.

## 2    Linear stability analysis and choice of liquids

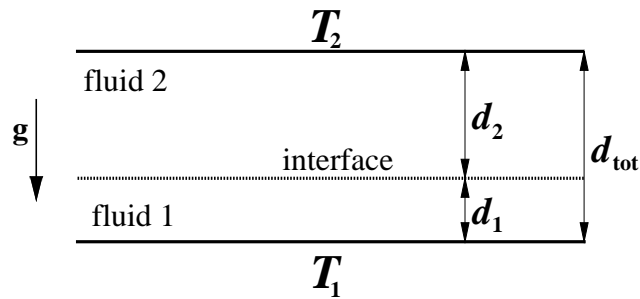

Fig. 2. Schematic diagram of the two-layer system. It consists of two superposed layers of immiscible fluids of respective depths $d_1$ and $d_2$. A vertical temperature difference $\Delta T = T_1 - T_2$ is applied either parallel or anti-parallel to the acceleration of gravity.

The theoretical model consists of two superposed layers of immiscible fluids. The layers have depths $d_1$ and $d_2$, where the subscripts 1 and 2 refer to the bottom and top fluids respectively (see Fig. 2). The system is bounded in the vertical direction by isothermal plates and is infinite in the horizontal direction. A temperature difference $\Delta T = T_1 - T_2$ is applied either parallel or antiparallel to the acceleration of gravity, so that $\Delta T > 0$ when the bottom plate is hotter than the top plate and $\Delta T < 0$ when the bottom plate is colder

Table 1

The density $\rho$, kinematic viscosity $\nu$, thermal conductivity $k$, constant pressure specific heat $c_p$, coefficient of thermal expansion $\beta$, surface tension with air $\sigma$, and temperature coefficient of surface tension are given at 25°C ($^a$ 26.8°C, $^b$ 25.10°C). The physical properties of the Galden liquids (perfluorinated hydrocarbons) were obtained from Ausimont, Inc. and the silicone oils from Dow Corning, Inc. and Palmer and Berg [22]. The properties of the remaining liquids are from ref. [23]. The units for kinematic viscosities are centiStokes (1 cS= $10^{-4}$m$^2$s$^{-1}$).

| Liquid | Galden HT-70 | Galden HT-100 | Galden HT-135 | acetonitrile | n-hexane |
|---|---|---|---|---|---|
| $\rho(10^3$ kg m$^{-3})$ | 1.68 | 1.78 | 1.73 | 0.776 | 0.655 |
| $\nu(10^{-4}$ m$^2$ s$^{-1})$ | 0.50 | 0.80 | 1.00 | 0.476 | 0.458 |
| $k(10^{-1}$ J m$^{-1}$ s$^{-1}$ K$^{-1})$ | 0.70 | 0.70 | 0.70 | 1.88 | 1.20 |
| $c_p(10^3$ J kg$^{-1}$ K$^{-1})$ | 0.962 | 0.962 | 0.962 | 2.23 | 2.27 |
| $\beta(10^{-3}$ K$^{-1})$ | 1.1 | 1.1 | 1.1 | 1.41 | 1.41 |
| $\sigma(10^{-3}$ N m$^{-1})$ | 14 | 15 | 17 | 28.66 | 17.89 |
| $\sigma_T(10^{-5}$ N m$^{-1}$ K$^{-1})$ | N/A | N/A | N/A | 12.63 | 10.22 |

| Liquid | water | silicone oil: 0.65 cS | silicone oil: 2 cS | silicone oil: 5 cS | silicone oil: 10 cS |
|---|---|---|---|---|---|
| $\rho(10^3$ kg m$^{-3})$ | 0.997 | 0.761 | 0.873 | 0.920 | 0.940 |
| $\nu(10^{-4}$ m$^2$ s$^{-1})$ | 0.893 | 0.65 | 2.00 | 5.00 | 10.00 |
| $k(10^{-1}$ J m$^{-1}$ s$^{-1}$ K$^{-1})$ | 6.09$^a$ | 1.004 | 1.088 | 1.172 | 1.339 |
| $c_p(10^3$ J kg$^{-1}$ K$^{-1})$ | 4.180 | 1.922$^b$ | 1.713 | N/A | 1.498 |
| $\beta(10^{-3}$ K$^{-1})$ | 0.257 | 1.34 | 1.17 | 1.05 | 1.08 |
| $\sigma(10^{-3}$ N m$^{-1})$ | 72.0 | 15.9 | 18.7 | 19.7 | 20.1 |
| $\sigma_T(10^{-5}$ N m$^{-1}$ K$^{-1})$ | 16.1 | N/A | N/A | N/A | 6.97 |

than the top plate. We refer to Rasenat *et al.* (15) for the governing equations. Linearization with respect to either stationary or oscillatory disturbances leads to a $12 \times 12$ matrix, whose determinant must vanish in order to yield a non-trivial solution. This condition leads to the desired dispersion relation. The results presented below are calculated with the numerical codes of Engel and Swift (20), who assumed a non-deformable interface, and VanHook (21), who considered the more general case with a deformable interface.

The predicted behavior depends critically on the physical properties of the liquids. We considered a large number of possible combinations and found a few that have properties yielding predictions for novel dynamics. The choice

Table 2
Fluid combinations studied in the two liquid layer experiments.

| Lower liquid | Upper liquid | Figure |
| --- | --- | --- |
| Galden HT-135 | 2 cS silicone oil | Fig. 7(b) |
| Galden HT-135 | 5 cS silicone oil | Fig. 7(a) and 6 |
| Galden HT-135 | 10 cS silicone oil | Fig. 17 |
| Galden HT-100 | 0.65 cS silicone oil | Fig. 10 and 11 |
| Galden HT-70 | 5 cS silicone oil | Fig. 12, 13 and 16 |
| Galden HT-70 | water | Fig. 9 |
| Acetonitrile | n-hexane | Fig. 14 and 15 |

of liquids was restricted by practical considerations that we discuss in Section 3.

We focus on three different classes of liquid combinations: perfluorinated hydrocarbon and silicone oil, perfluorinated hydrocarbon and water, and acetonitrile and n-hexane. The material properties relevant to Bénard-Marangoni convection are listed in Table 1 and the detailed list of combinations is given in Table 2. The perfluorinated hydrocarbons are manufactured by Ausimont Inc.[4] and are called Galden fluids, the silicone oils are produced by Dow Corning, Inc.[5], and the acetonitrile and n-hexane are chromatography grades from EM Science. The interfacial tensions and their temperature coefficients are not known for these liquid combinations. However, they can be estimated using Antonow's rule (24), which states that the interfacial tension between two liquids is equal to the difference between their surface tensions in air. Thus we include the surface tension and its temperature coefficient for each liquid, when available.

The fraction of the total forcing due to buoyancy can be estimated by calculating the parameter $\alpha$,

$$\alpha = (1 + \frac{M_c}{M_0}\frac{R_0}{R_c})^{-1}, \qquad (1)$$

introduced by Parmentier *et al.* (25), where $M_c = (\sigma_T d_1 \Delta T_c)/(\eta_1 \kappa_1)$ is the critical Marangoni number of the full two layer problem, $R_c = (g\beta_1 d_1^3 \Delta T_c)/(\nu_1 \kappa_1)$ is the critical Rayleigh number and $\Delta T_c$, $\sigma_T$, $\eta_1$, $\nu_1$, $\beta_1$, $\kappa_1$ and $g$ are, respectively, the critical temperature difference applied across both layers, the temperature coefficient of interfacial tension, the dynamic and kinematic viscosities, the coefficient of thermal expansion, the thermal diffusivity and the acceleration of gravity. The suffix 1 refers to the liquid in the bottom layer. $R_0$ is the critical Rayleigh number for the onset of convection in the absence of Marangoni forces, and $M_0$ is the critical Marangoni number in the absence of buoyancy forces. $R_0$ and $M_0$ are both determined from linear stability calculations for each of the studied systems.

---
[4] http:\\www.ausimont.com
[5] http:\\www.dowcorning.com

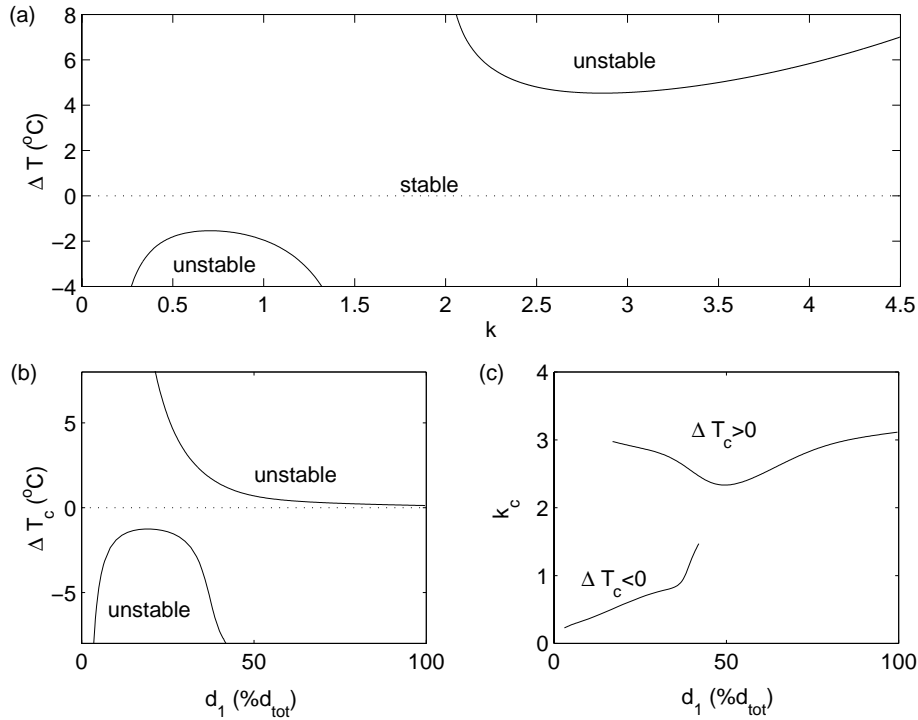

Fig. 3. (a) Marginal stability curves for HT-70 and 5 cS silicone oil with $d_1 = 0.27\,d_{\text{tot}}$, $d_{\text{tot}} = 3.00$ mm, and $\sigma_T = 7.3 \times 10^{-5}$ N m$^{-1}$ K$^{-1}$ ($\alpha = 0.48$). $k$ is the horizontal wavenumber in dimensionless form, $k = 2\pi d_1/\lambda$, where $\lambda$ is the wavelength of the pattern. Note that $\Delta T > 0$ corresponds to heating from below and $\Delta T < 0$, heating from above. (b) Dependence of the critical temperature difference on the bottom depth for $d_{\text{tot}} = 3.00$ mm and $\sigma_T = 7.3 \times 10^{-5}$ N m$^{-1}$K$^{-1}$. (c) Dependence of the critical horizontal wavenumber $k_c$ on $d_1$.

The onset of convection heating from above ($\Delta T < 0$) is predicted in all three classes of liquid combinations for specific depth ratios. As an example, we first present results for the instability in the system of Galden HT-70 and 5 cS silicone oil where the total depth is $d_{\text{tot}} = 3.00$ mm. The marginal stability of the system when $d_1 = 0.27\,d_{\text{tot}}$ ($\alpha = 0.48$) is presented in Fig. 3(a). The curves delimit regions of stationary instabilities of critical horizontal wavenumbers, $k_{c(\text{th})} = 0.706$ and $k_{c(\text{th})} = 2.798$ for $\Delta T < 0$ and $\Delta T > 0$ respectively. The wavenumbers are given in dimensionless form, $k = 2\pi d_1/\lambda$, where $\lambda$ is the wavelength of the pattern. (Similarly, oscillation frequencies $\omega$ will be given in dimensionless form, $\omega = (2\pi/\tau)/(d_1^2/\kappa_1)$, where $\tau$ is the period of the oscillation.) Figure 3(c) suggests that the thicker upper layer imposes the wavelength when $\Delta T < 0$ and *vice versa* when $\Delta T > 0$, in accordance with the mechanisms illustrated by Fig. 1. In Fig. 3(b), the critical temperature difference for the onset of convection is plotted as a function of the fractional bottom depth. It is clear that the instability heating from above ($\Delta T < 0$) is most readily observed when $d_1$ is around 20% of the total depth of the system.

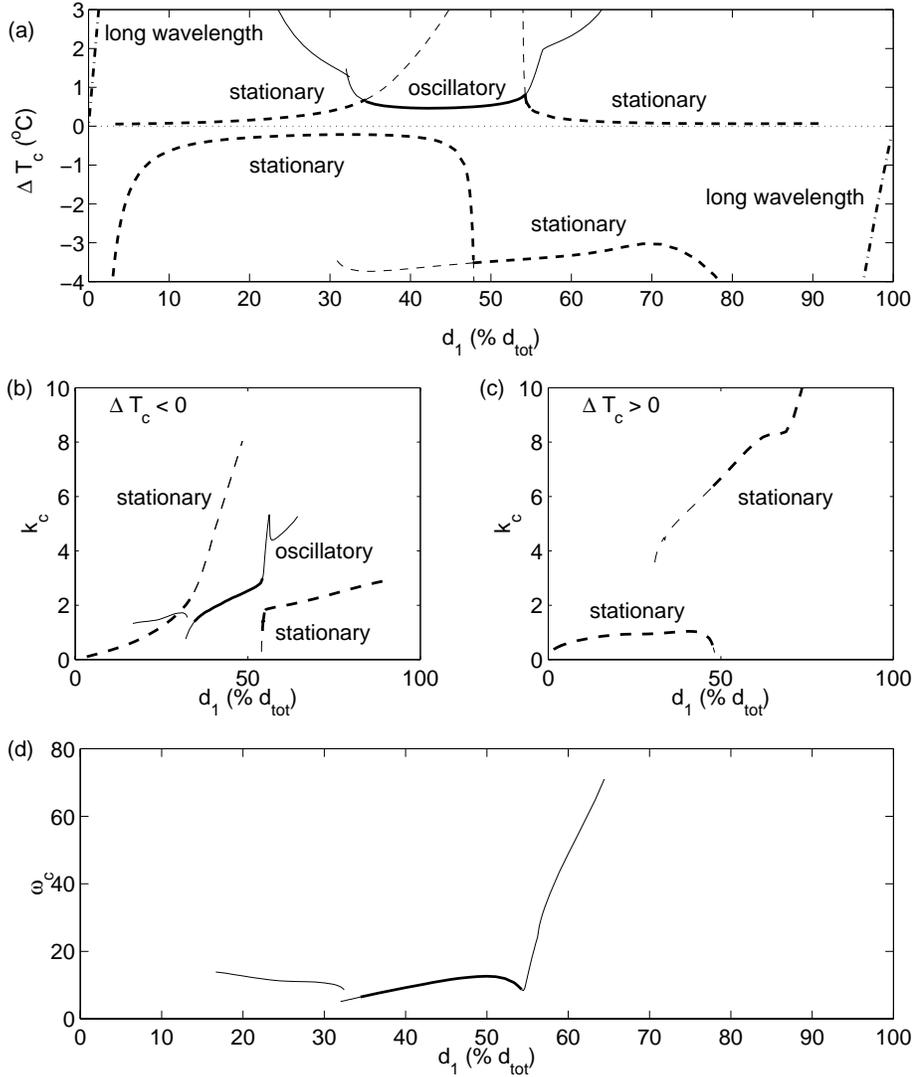

Fig. 4. (a) Dependence of the critical temperature difference on the depth of the bottom layer for $d_{\text{tot}} = 4.50$ mm and $\sigma_T = 1.0 \times 10^{-4}$ N m$^{-1}$ K$^{-1}$. The stationary (oscillatory) bifurcation branches are plotted with dashed (solid) lines. The instabilities encountered in the physical system are drawn with bold lines. Long wavelength instabilities dominate for $\Delta T > 0$ ($\Delta T < 0$) when the lower (upper) layer is very thin. For $\Delta T > 0$, there are two distinct stationary instabilities separated by an oscillatory domain, where the bifurcation branches cross at codimension-2 points. For $\Delta T < 0$, two different stationary instabilities are observed. (b) Dependence of the critical horizontal wavenumber $k_c$ on $d_1$ when $\Delta T_c > 0$. (c) Dependence of the critical horizontal wavenumber $k_c$ on $d_1$ when $\Delta T_c < 0$; separate branches associated with the various instabilities point to differences in the underlying fluid dynamics. (d) Dependence of the critical frequency $\omega_c$ of the oscillation on $d_1$ ($\omega_c$ is given in non-dimensional form, $\omega_c = (2\pi/\tau_c)(d_1^2/\kappa_1)$ where $\tau_c$ is the critical period of the oscillation.)

The onset of oscillatory convection is predicted in the acetonitrile and n-hexane system but not in the perfluorinated hydrocarbon and water or silicone oil systems for the total depths we examined ($d_{\text{tot}} < 4.60$ mm). As an example, we focus on an acetonitrile and n-hexane system with a total depth of 4.50 mm and we set the temperature coefficient of interfacial tension to $\sigma_T = 1.0 \times 10^{-4}$ N m$^{-1}$ K$^{-1}$. This value was chosen since it leads to satisfactory agreement with our experimental results that we will describe in Section 4.3. The dependence of the critical temperature difference on the fractional bottom depth is plotted in Fig. 4(a), and the corresponding dependences of the critical wavenumber and frequency are shown in Figs. 4(b),(c) and (d).

The rich bifurcation structure of this configuration is exposed in Fig. 4(a). Long wavelength instabilities are predicted when the lower (upper) layer is very thin (on the order of 1% of the total depth) for $\Delta T > 0$ ($\Delta T < 0$). Thus in both cases, Marangoni forces destabilize the shallow layer, in accordance with the mechanisms illustrated in Fig. 1. The system also exhibits four different short-wavelength stationary bifurcation branches, two of which are found when heating from above. The contrast between their wavenumbers points to qualitatively different underlying fluid dynamics. Most interestingly, the stationary branches that exist for $\Delta T > 0$ are separated by an oscillatory region (when $d_1$ is around 45% of the total depth). Two crossing oscillatory branches are displayed in Fig. 4(a), but only one of them would be observed physically for this set of parameters. Interesting dynamics are expected close to the codimension-2 points where stationary and oscillatory bifurcation branches cross.

The dependence of the dynamics of the acetonitrile and n-hexane system on the total depth is illustrated in Fig. 5, where the critical temperature difference for the onset of convection is plotted against the fractional bottom depth, for total depths of 1.5 mm, 3.0 mm, 4.5 mm and 6.0 mm. The corresponding wavenumbers and the frequencies of the oscillatory instabilities are shown in Fig. 6; for clarity, the long wavenumber instabilities are not included. The oscillatory instability is present for all four depths but its critical onset temperature difference strongly increases as the total depth of the system is reduced and so do the critical temperature differences for the onset of stationary convection heating from below. The critical wavenumbers only weakly depend on the total depth, whereas the critical frequency of the oscillatory instability increases as $d_{\text{tot}}$ is reduced. For heating from above, the critical parameters hardly vary with the total depth of the system, when $d_1$ is between 5% and 40% of the total depth of the system. In addition, when $d_{\text{tot}}$ is reduced from 3 mm to 1.5 mm the two disconnected bifurcation branches observed for $\Delta T < 0$ merge to form a single branch, whose corresponding wavenumber is continuous across the range of depth ratios.

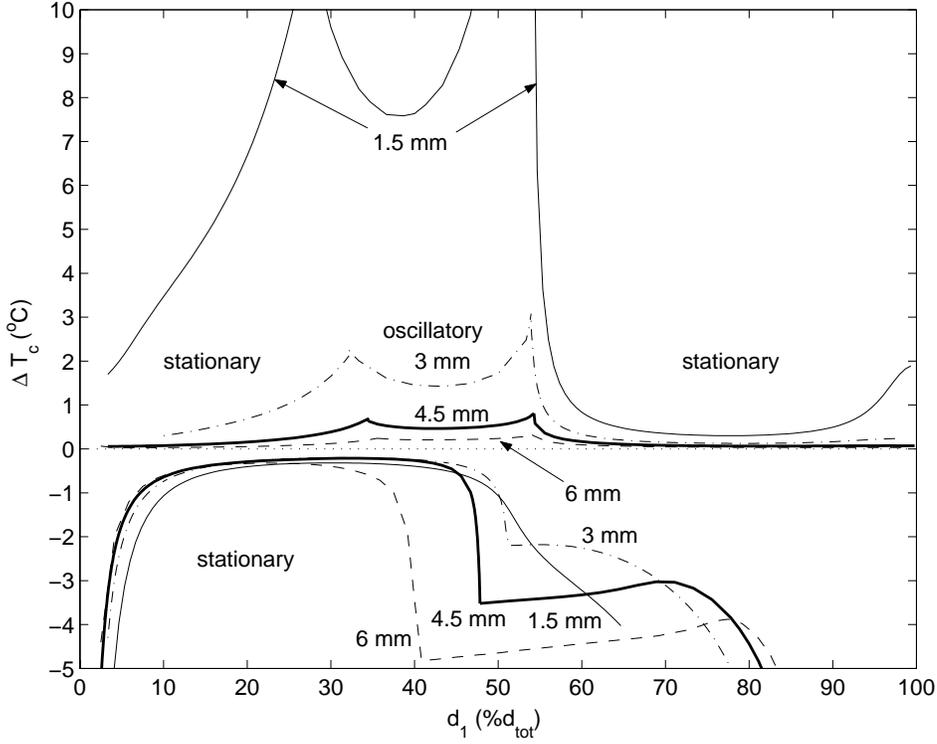

Fig. 5. Dependence of the critical temperature difference for the onset of convection in the acetonitrile and n-hexane on the fractional bottom depth. The bifurcation structures in systems of total depths of $d_{\text{tot}} = 1.5$ mm, $d_{\text{tot}} = 3.0$ mm, $d_{\text{tot}} = 4.5$ mm and $d_{\text{tot}} = 6.0$ mm are plotted with solid, dot-dashed, bold solid and dashed lines, respectively. When $\Delta T > 0$, the critical temperature difference strongly increases as $d_{\text{tot}}$ is reduced. However, the instability heating from above depends only weakly on the total depth of the system when $d_1$ is between 0.05 and 0.40 $d_{\text{tot}}$. Also, as the total depth is reduced from 3 mm to 1.5 mm, the two disconnected bifurcation branches observed for $\Delta T < 0$ merge.

## 3 Experimental apparatus

A gold-plated copper mirror and a sapphire window, both 1.27 cm thick and 10.16 cm in diameter, form the bottom and top plates of the convection cell, respectively. A spacer ring serves as a sidewall and also sets the total depth of the system. For the Galden and silicone oil experiments, the boundary was made of Teflon, which is preferentially wetted by the perfluorinated liquid, to achieve configurations with a very thin lower layer. In the absence of Teflon sidewalls, it is not energetically favorable for thin layers of perfluorinated liquid to cover the entire bottom plate when the lighter liquid is silicone oil. Fiber glass epoxy composite (G10) sidewalls were used for the other combinations. For acetonitrile and n-hexane, the cell was sealed with Teflon encapsulated o-rings, as common o-ring materials are not resistant to both solvents. Thus, the convection cell is cylindrical and has an inner diameter between 8.25 cm and 8.56 cm depending on the spacer ring used. The parallelism of the top and bottom plates was adjusted interferometrically to within $\pm 5\,\mu$m across the diameter for G10 boundaries and $\pm 10\,\mu$m for Teflon spacers. The level of the convection cell was set with a bubble level to within $1.1 \times 10^{-3}$ rad.

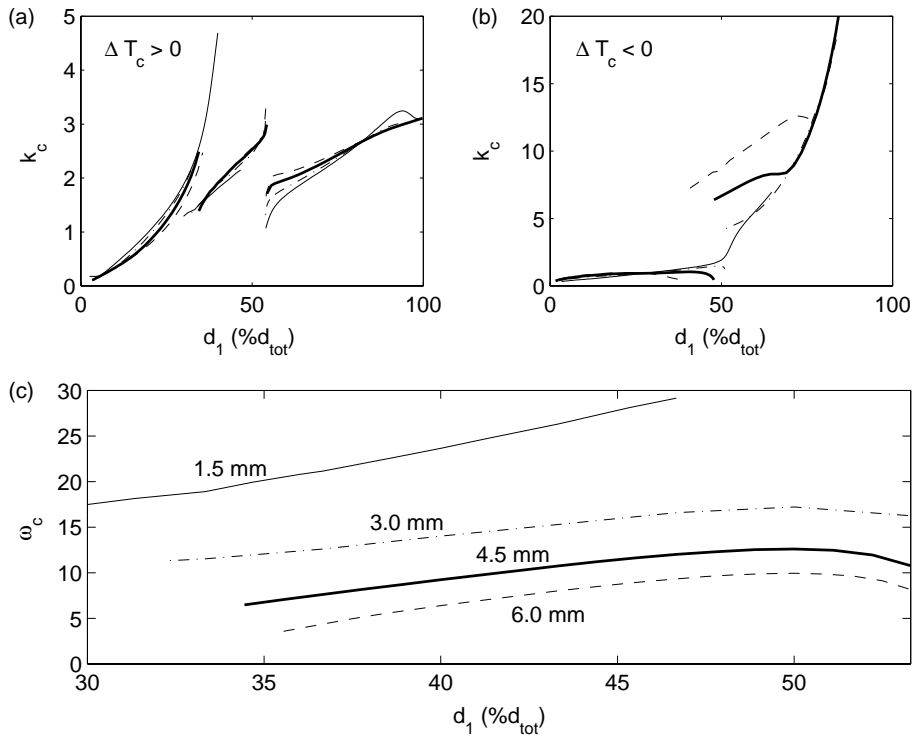

Fig. 6. (a) Dependence of the critical horizontal wavenumber $k_c$ on the fractional bottom depth when $\Delta T > 0$ for the system shown in Fig. 5. (b) Dependence of the critical horizontal wavenumber $k_c$ on the fractional bottom depth when $\Delta T < 0$. (c) Dependence of the critical frequency $\omega_c$ associated with the oscillatory instabilities shown in Fig. 5. $\omega_c$ increases substantially when $d_{tot}$ is reduced from 6.0 mm to 1.5 mm.

Each of the top and bottom surfaces of the cell is kept isothermal within 0.005°C over periods of hours by immersion in independent turbulent water baths. Thus, a vertical temperature gradient may be applied by heating from above and cooling from below or *vice versa*. The mean temperature of the system was typically kept constant within 0.01°C during the experimental runs. Temperature was measured on the top and bottom surfaces of the convection cell with thermistors. Measurements with a movable probe showed that the uniformity of the temperature along a diameter of the sapphire window was better than 0.005°C.

The layering of the liquids is a delicate procedure, whose success was found to depend on the values of the interfacial tension, the viscosities, and most importantly the wetting properties within the cell. The cell was first filled entirely with the denser liquid, which was in turn partially removed as the lighter liquid was slowly injected above the rim of the sidewall using a syringe pump. The layering of most sets of liquids was metastable, so that the layers could not be recovered after tilting the cell. Thus, the depth ratio was determined destructively after each experiment, based on the location of the interface when the cell was placed in a vertical position.

The convection patterns were visualized with a standard shadowgraph system. In Bénard-Marangoni convection, superposed convection cells in the two layers are typically counter-rotating (20) (viscous coupling), and the intensity of the image corresponds to the differential between the convective motion in each layer. The optical properties of the liquids may weaken the shadowgraph image and meniscus effects may create non-uniformities in the thickness of the layers. Also, although meniscus effects could distort the outer edge of the picture considerably, the central region would remain unperturbed.

To allow for a well-defined interface, we chose pairs of immiscible liquids for the experiments. Unfortunately, low solubility and low interfacial tension are mutually exclusive criteria. As will be discussed in Section 4.1, thermo-solutal effects were observed in the lowest viscosity grades of the Galden and silicone oil combination, for which substantial mixing took place. However, even in nominally immiscible liquids, such as the Galden and water combination (water is soluble in Galden to 14 ppm against 38 ppm for air), thermally-driven migration of the lighter species through the bottom layer to the surface of the mirror or *vice versa* was observed. This impaired the experimental measurements, as will be discussed in Section 4.4.

## 4 Experimental results

### 4.1 Heating from below: stationary convection

A variety of convection patterns were found for $\Delta T > 0$ for the Galden and silicone oil system and the Galden and water system, in both cases in qualitative agreement with linear stability predictions. In most of the Galden and silicone oil configurations, for total depths ranging from $d_{\text{tot}} = 1.89$ mm to $d_{\text{tot}} = 4.60$ mm and a variety of depth ratios $(d_2/d_1)$ ranging from 0.62 to 2.8, hexagons were observed at onset, as illustrated in Fig. 7(a), and a secondary transition to a mixed state of squares and hexagons took place well above onset, as shown in Fig. 7(b). A background image was subtracted from each of these pictures to improve the signal to noise ratio. These are analogous to the traditional Marangoni patterns observed in liquid-gas layers. A defect-free square planform could not be reached because mixing phenomena occurred for large values of the applied temperature difference, as discussed in Section 4.4.

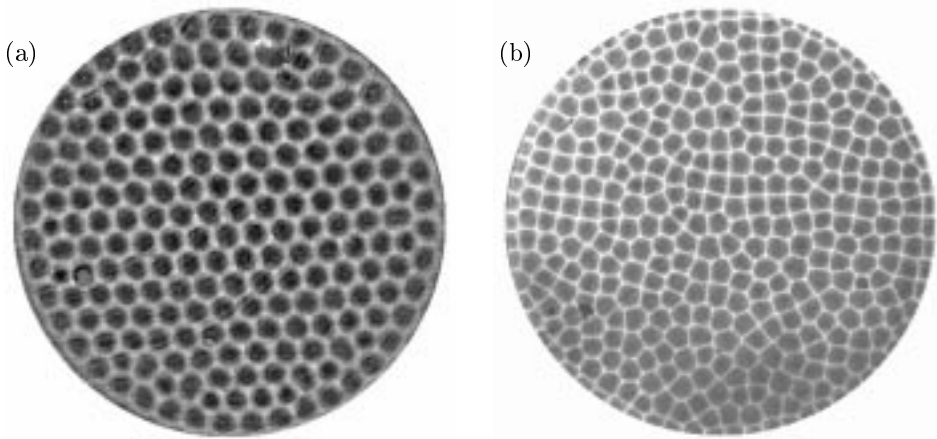

Fig. 7. Convection patterns observed in Galden and silicone oil heated from below: (a) hexagonal pattern near onset in Galden HT-135 and 5 cS silicone oil ($\Delta T = 0.960°$C, $d_{\text{tot}} = 4.60$ mm, $d_1 = 0.415\, d_{\text{tot}}$, $\alpha = 0.60$); (b) mixed state of squares and hexagons far above onset in Galden HT-135 and 2 cS silicone oil ($\Delta T = 2.805°$C, $d_{\text{tot}} = 1.89$ mm, $d_1 \approx 0.54\, d_{\text{tot}}$, $\alpha = 0.15$).

In addition, we studied the onset of hexagonal patterns in Galden HT-135 and 5 cS silicone oil for $d_{\text{tot}} = 4.35$ mm and $d_1 = 0.493\, d_{\text{tot}}$ ($\alpha = 0.60$); the difficulties reported in Section 4.4 did not significantly affect this particular system. Additional temperature control was provided by PID controlled heaters on the water lines leading into each of the isothermal baths. As a result, the standard deviation of the applied temperature difference fluctuations was reduced to $5 \times 10^{-4}°$C in these experiments. The onset of convection was measured by increasing the applied temperature difference in steps of $0.02°$C from the conductive state into the convective regime. For each applied temperature difference, the system was left to settle for 32 vertical diffusion times ($t_v = d_1^2/\kappa_1$, where $\kappa_1$ is the thermal diffusivity of the lower liquid) before a shadowgraph image was sampled. This procedure was repeated for decreasing temperature differences and in a second run, waiting times were raised to $96 t_v$ in order to check reproducibility. Fast Fourier Transforms of the central part of each image yielded the power of the pattern and its wavenumber. The wavenumber was found to remain constant near onset and equal to $k_{c(\text{exp})} = 2.50 \pm 0.05$, close to the theoretical value of $k_{c(\text{th})} = 2.351$.

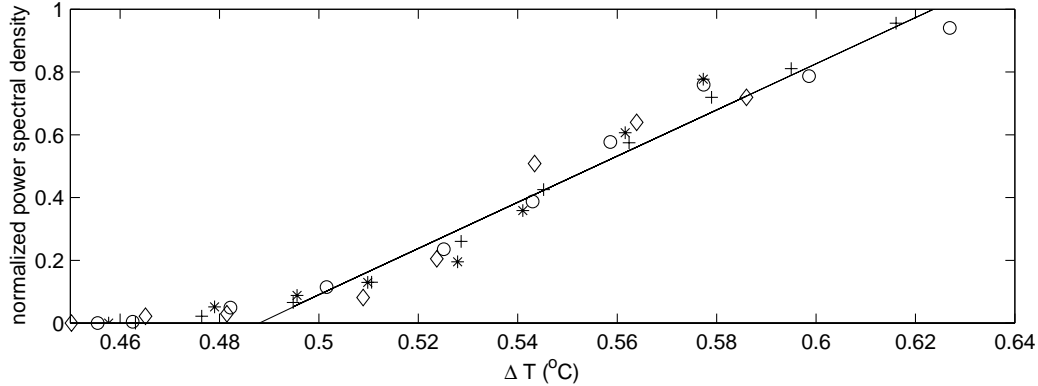

Fig. 8. Normalized power spectral density of the hexagonal convection pattern in Galden HT-135 and 5 cS silicone oil shown in Fig. 7(a): ○ increasing $\Delta T$, waiting for $32 t_v$; + decreasing $\Delta T$, waiting for $32 t_v$; ◇ increasing $\Delta T$, waiting for $96 t_v$; ∗ decreasing $\Delta T$, waiting for $96 t_v$; — linear fit to the experimental data. ($t_v$ is the vertical diffusion time in the lower layer, $t_v = d_1^2/\kappa_1$.)

The normalized power of the hexagonal pattern is plotted against the applied temperature difference in Fig. 8. The data obtained when leaving the system to settle for $32 t_v$ or $96 t_v$ are in good agreement. The transition from conduction to convection takes place through an imperfect bifurcation. In careful experiments in a thin layer of silicone oil with a free surface, Schatz et al. (3) uncovered a subcritical onset of convection, as predicted by the weakly nonlinear theory. However, the hysteresis was very sensitive to perturbations and faded as the onset was repeatedly monitored.

The critical temperature difference, obtained by linear fit to the data in Fig. 8, is $\Delta T_{c(exp)} = 0.488 \pm 0.005 °C$. Experiments and linear theory are in agreement when the temperature coefficient of interfacial tension is set to $\sigma_T = (7.3 \pm 0.2) \times 10^{-5}$ N m$^{-1}$ K$^{-1}$. This value is further used as an estimate for all the perfluorinated hydrocarbon and silicone oil systems.

For the Galden HT-70 and water combination, the depth was set to $d_{tot} = 3.24$ mm and several depth ratios were investigated. The onset of convection led to a square planform, whereas a secondary transition led to rolls, as in the set of liquids (FC-75 and water) discussed by (10). Shadowgraph images of both convective states are shown in Fig. 9. A background image has been subtracted from the picture of squares shown in Fig. 9(a). The critical temperature difference for the onset of squares for $d_1 = 0.315 \, d_{tot}$ ($\alpha = 0.49$) was estimated by increasing the applied temperature difference through onset in steps of $0.02°C$ and waiting for $150 t_v$ at each step. By taking the power of the pattern in the central part of each image as described above, a critical temperature difference of $\Delta T_{c(exp)} = 1.476 \pm 0.005 °C$ was determined. In order to obtain this value theoretically, the temperature coefficient of interfacial tension was adjusted to $\sigma_T = (4.10 \pm 0.06) \times 10^{-5}$ N m$^{-1}$ K$^{-1}$. The experimen-tal dimensionless wavenumber at onset $k_{c(\text{exp})} = 2.71 \pm 0.05$ was close to the corresponding theoretical prediction, $k_{c(\text{th})} = 2.394$. Although Engel and Swift (20) predict the subcritical onset of hexagons, they also find a transition to squares very close above onset, and conclude that it is conceivable that only squares are observed in practice.

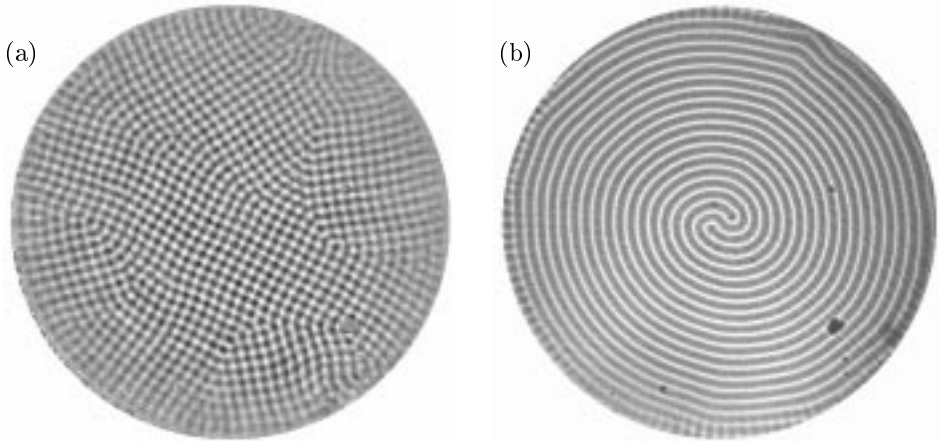

Fig. 9. Convection patterns in Galden HT-70 and water heated from below for $d_1 = 0.315\, d_{\text{tot}}$, $d_{\text{tot}} = 3.24$ mm: (a) square planform close to onset, $\Delta T = 2.110°$C, $\epsilon = (\Delta T - \Delta T_c)/\Delta T_c = 0.43$; (b) four-armed spiral, $\Delta T = 2.823°$C, $\epsilon = 0.91$. The shadowgraph picture was cropped to remove its outer edge, distorted by meniscus effects.

For larger temperature differences, the secondary transition to rolls would typically yield a 'Pan-Am' pattern where the rolls were forced to arrange perpendicularly to the sidewalls at the boundary. However, forcing due to the presence of a meniscus at the sidewalls would in some instances lead to parallel rolls at the boundary. The pattern then took the form of a spiral, as shown in Fig. 9(b), whose core slowly rotated clockwise at a rate of approximately 25 deg/hour. The spiral pattern became unstable to cross-rolls originating at the side-walls for large values of $\Delta T$.

Although the perfluorinated hydrocarbons are rated insoluble with silicone oil by the manufacturer, substantial mixing occurs with the lowest viscosity grade of silicone oil (0.65 cS). Experiments with Galden HT-100 in the bottom layer were run in a shallow cell of total depth of $d_{\text{tot}} = 1.47$ mm, which could be achieved due to the very low interfacial tension of this system. Two depth ratios were considered, $d_1 = 0.70\, d_{\text{tot}}$ and $d_1 = 0.26\, d_{\text{tot}}$. In both cases, convective motion was observed significantly below the predicted onset temperature differences. For $d_1 = 0.70\, d_{\text{tot}}$, a faint square planform was observed at 49% of the critical temperature difference predicted theoretically; larger temperature differences led to a transition to a mixed state of squares and hexagons. The temperature difference was incremented in steps of 0.02°C and the system was then left to settle for at least $504 t_v$. The convection patterns observed for $d_1 = 0.26\, d_{\text{tot}}$ are displayed in Fig. 10 for increasing applied temperature differences. When the system was left isothermal, a large number of droplets of silicone oil would spontaneously wet the mirror and thus form bridges between the upper layer and the bottom plate, as shown in Fig. 11.

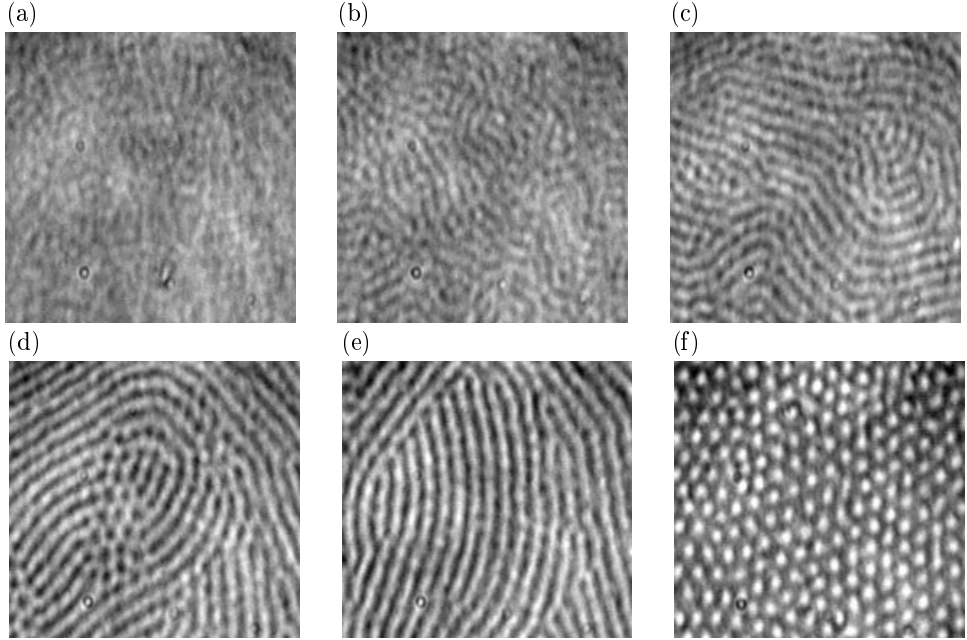

Fig. 10. Convection patterns in Galden HT-100 and 0.65 cS silicone oil heated from below ($d_{\text{tot}} = 1.47$ mm, $d_1 = 0.26\, d_{\text{tot}}$). Each image corresponds to the 3.74 cm × 3.74 cm central part of the convection cell: (a) $\Delta T = 0.33°$C, (b) $\Delta T = 0.89°$C, (c) $\Delta T = 1.27°$C, (d) $\Delta T = 2.43°$C, (e) $\Delta T = 2.81°$C, (f) $\Delta T = 3.40°$C. Stability theory predicts the onset of convection (with an hexagonal planform) for $\Delta T_{c(th)} = 5.497°$C, assuming $\sigma_T = 7.3 \times 10^{-5}$ N m$^{-1}$ K$^{-1}$.

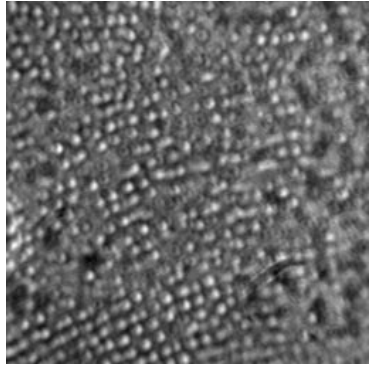

Fig. 11. Shadowgraph image of the central region (3.74 cm × 3.74 cm) of the convection cell filled with Galden HT-100 and 0.65 cS silicone oil (same setup as shown in Fig. 10) in a near isothermal state ($\Delta T = -0.03°$C). The bright dots correspond to droplets of silicone oil in contact with the bottom plate, indicating that the interface has a bridged configuration.

This interfacial configuration disappeared as soon the system was heated from below. Weak intermingled rolls then appeared, and their characteristic length scale increased as the applied temperature difference was raised, as shown in Fig. 10(a),(b),(c). In Fig. 10(d), some dark-centered hexagons can be observed, but they do not appear in Fig. 10(e), where the planform consists of almost straight rolls. Finally, a gradual transition to light-centered hexagons takes place as shown in Fig. 10(f). The theoretical temperature difference for the onset of convection was calculated to be $\Delta T_{c(th)} = 5.497°C$, far above the values reported in Fig. 10. Thus, these peculiar patterns, whose wavelengths are more than a factor of two larger than the Bénard-Marangoni predictions, suggest that destabilizing thermosolutal gradients in the system may drive the initial onset of convection patterns (26).

*4.2 Heating from above*

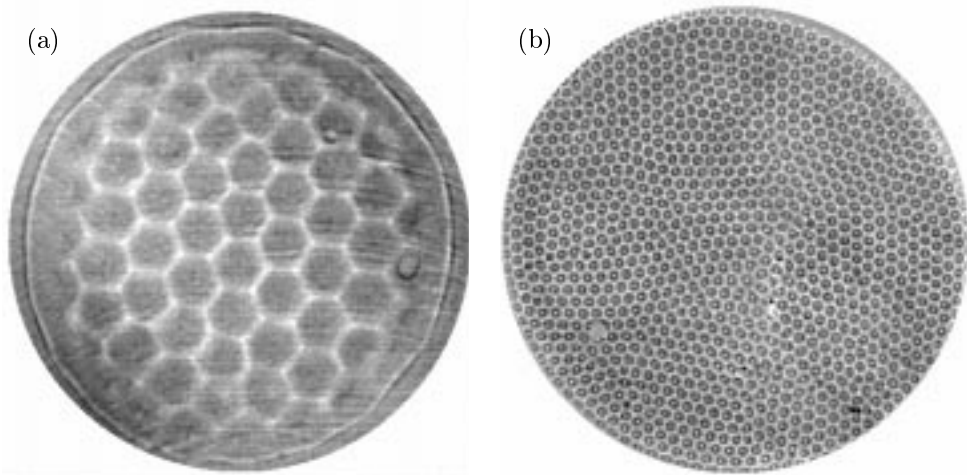

Fig. 12. Comparison of convection close to onset for heating from above and below in Galden HT-70 and 5 cS silicone oil ($d_{\text{tot}} = 3.06$ mm): (a) $\Delta T = -3.778°C$ and $d_1 = 0.26\, d_{\text{tot}}$ ($\Delta T_{c(th)} = -1.478°C$); (b) $\Delta T = 3.295°C$ and $d_1 = 0.31\, d_{\text{tot}}$ ($\Delta T_{c(th)} = 2.894°C$). Marangoni forces are destabilizing in the thicker upper (thin lower) layer when the heating is applied from above (below), and it appears that the wavelength of the pattern is dominantly influenced by the depth of the destabilized layer.

Onset of convection for $\Delta T < 0$ was observed in both the perfluorinated hydrocarbon and silicone oil and the acetonitrile and n-hexane systems, but not in the perfluorinated hydrocarbon and water system, although it was predicted by linear stability calculations in all three systems for specific depth ratios. The perfluorinated hydrocarbon and silicone oil system is comprised of Galden HT-70 and 5 cS silicone oil with depth $d_{\text{tot}} = 3.06$ mm. The experiments were very delicate because convection for $\Delta T < 0$ is only predicted within a narrow range of depth ratios accessible experimentally, as shown in Fig. 3. In addi-tion, droplets of silicone oil condensed on the mirror and thus obscured the visualization system within hours, as will be discussed in Section 4.4. Background subtracted shadowgraph images of convection patterns observed for $d_1 = 0.26\,d_{\text{tot}}$ and $\Delta T = -3.778°$C and for $d_1 = 0.31\,d_{\text{tot}}$ ($\alpha = 0.48$) and $\Delta T = 3.295°$C are shown in Fig. 12(a) and (b) respectively. Both patterns are hexagonal in agreement with the predictions of Engel and Swift (20). The wavelength of the convection pattern for $\Delta T < 0$ shown in Fig. 12(a) is larger than that observed for $\Delta T > 0$ in the same system by a factor of 4.5. The dimensionless wavenumbers of the patterns shown in Fig. 12 were calculated to be $k_{c(\text{exp})} = 0.60$ and $k_{c(\text{exp})} = 2.75$ respectively, in satisfactory agreement with the theoretical predictions of $k_{c(\text{th})} = 0.693$ and $k_{c(\text{th})} = 2.789$.

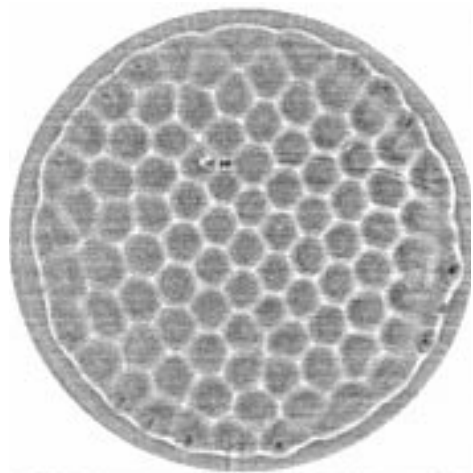

Fig. 13. Hexagonal convection pattern in Galden HT-70 and 5 cS silicone oil for $\Delta T = -3.485°$C, $d_1 = 0.20\,d_{\text{tot}}$ and $d_{\text{tot}} = 3.06$ mm ($\Delta T_{c(th)} = -1.242°$C). The wavenumber of the pattern is $k_{\text{exp}} = 0.61$ in satisfactory agreement with theoretical predictions of $k_{c(\text{th})} = 0.573$. The dimensional wavelength of this pattern is shorter than that shown in Fig. 12(a) by a factor of 1.28; the theoretical prediction is a factor of 1.52.

In Fig. 13, we show an additional hexagonal convection pattern observed in Galden HT-70 and 5 cS silicone oil for $d_1 = 0.20\,d_{\text{tot}}$ and $d_{\text{tot}} = 3.06$ mm when $\Delta T < 0$. The dimensionless experimental wavenumber is $k_{c(\text{exp})} = 0.61$ and linear theory calculations yield $k_{c(\text{th})} = 0.573$.

4.3 *Heating from below: oscillatory convection*

Oscillatory convection was found in the acetonitrile and n-hexane system for $d_{\text{tot}} = 4.54$ mm and $d_1 = 0.39\,d_{\text{tot}}$ ($\alpha = 0.41$) when $\Delta T > 0$. The applied temperature difference was raised in 6 steps of 0.04°C and the system was left to settle for $43 t_{\text{v}}$ at each temperature before time series of shadowgraph images were sampled for 1000 s at a rate of 0.86 Hz. During the run, the picture became substantially obscured due to condensation of acetonitrile on the sapphire window, as discussed in Section 4.4. The signal-to-noise ratio was improved by recording background images at the end of a run and subtracting them from each sample. The first observed convection pattern was a time-periodic wave, which was initially localized in a small region close to the side boundary. As the temperature difference was increased, the system displayed a mixed state of wave fronts and steady cells, as can be seen in the snapshots of Fig. 14. At higher temperature differences, the flow became disordered and the sources of waves spread to the interior of the cell. The time evolution over 200 s of a segment of the image of a length of 2.66 cm and averaged over a width of 0.76 cm is shown in Fig. 15. The periodicity in the space-time plane indicates a standing wave pattern.

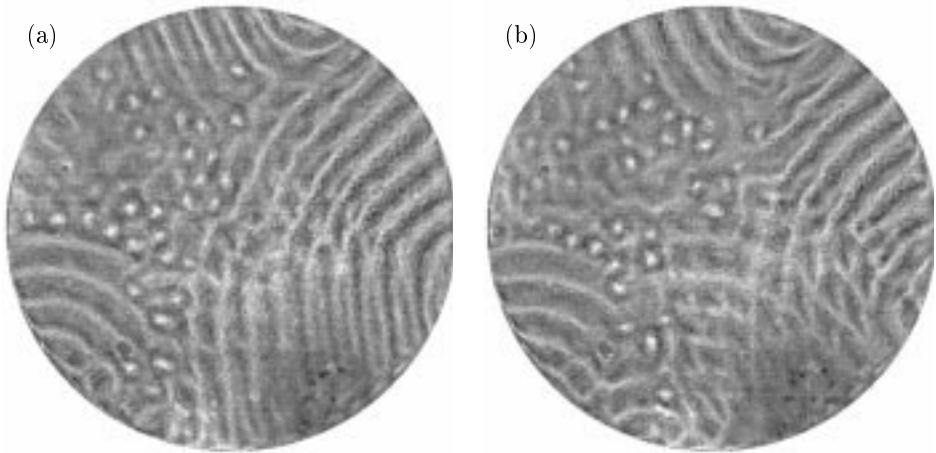

Fig. 14. Snapshots of oscillatory convection patterns in the acetonitrile and n-hexane system heated from below, from a time series recorded for $\epsilon = (\Delta T - \Delta T_{c(exp)})/\Delta T_c = 0.86$ ($\Delta T_{c(exp)} = 0.922°$C), $d_{\text{tot}} = 4.54$ mm and $d_1 = 0.39\, d_{\text{tot}}$. The wave fronts are travelling in different directions, forming rolls, squares and diamonds. (a) time = 0 s, (b) time = 232 s.

The wavenumber of the pattern was determined by computing the two-dimensional FFT on a central square region of the image (5.53 cm×5.53 cm). The power spectra of time series extracted for each pixel were also calculated, and then averaged in order to estimate the critical frequency $\omega_{c(exp)}$ of the standing wave. Both of these quantities remained constant as the temperature difference was increased past onset. The experimental values of $k_{c(exp)} = 2.00 \pm 0.05$ and $\omega_{c(exp)} = 11.5 \pm 0.7$ are in satisfactory agreement with linear stability calculations, which predict $k_{c(th)} = 1.84$ and $\omega_{c(th)} = 8.734$, when $\sigma_T = 1.0 \times 10^{-4}$ N m$^{-1}$ K$^{-1}$. The wavenumber did not vary as $\sigma_T$ was reduced to $3.0 \times 10^{-5}$ N m$^{-1}$ K$^{-1}$, close to the value predicted by Antonow's rule (24), but the frequency decreased to $\omega_{c(th)} = 2.46$, diverging from the value measured experimentally.

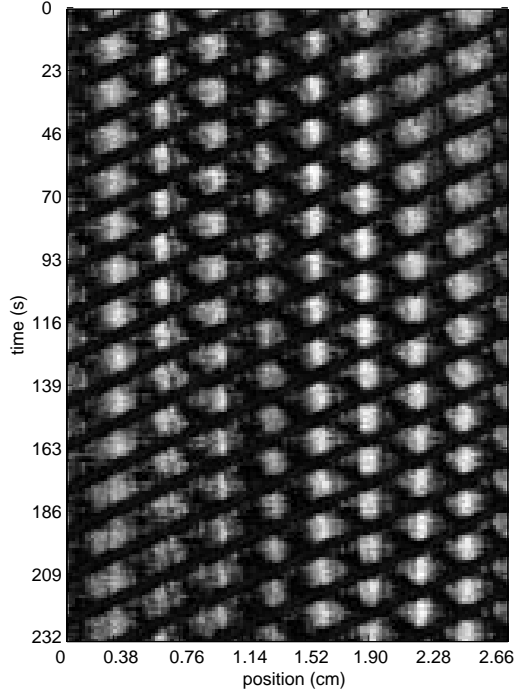

Fig. 15. Space-time diagram of the oscillatory pattern in the acetonitrile and n-hexane system heated from below, averaged over a 0.76 cm wide region of the convection cell ($\epsilon = 0.86$, $d_{\text{tot}} = 4.54$ mm and $d_1 = 0.39\, d_{\text{tot}}$). The periodicity of the intensity of the shadowgraph picture in the space-time plane indicates the presence of standing waves.

The critical temperature difference for the onset of convection was extrapolated from the power spectral density measurements of the patterns above onset and found to be $\Delta T_{c(exp)} = 0.686 \pm 0.012°$C. The theoretical critical temperature difference depends very weakly on the value of the temperature coefficient of the interfacial tension and varies from $\Delta T_{c(th)} = 0.470°$C to $0.503°$C, when $\sigma_T$ is decreased from $1.0 \times 10^{-4}$ N m$^{-1}$ K$^{-1}$ to $3.0 \times 10^{-5}$ N m$^{-1}$ K$^{-1}$. These values are both in satisfactory agreement with the measured value $\Delta T_{c(exp)}$. Hence, $\sigma_T$ could not be accurately determined from this experiment.

In addition, mixed stationary patterns of squares and rolls were observed in the acetonitrile and n-hexane system heated from below for $d_1 = 0.28 d_{tot}$, in good agreement with the linear predictions shown in Fig 4(a). Also, hexagonal patterns were seen for both depth ratios when the system was heated from above.

## 4.4 Mixing effects

In addition to the thermo-solutal-driven convection patterns observed in partially miscible combinations of Galden HT-100 and 0.65 cS silicone oil and described in Section 4.1, a more subtle effect of mixing, also dependent on the direction of the applied temperature difference, was observed to impair the study of convective motion. When $\Delta T > 0$, the denser species constituting the lower layer would migrate through the upper layer and condense onto the sapphire window, thus forming a thin, uniform film of droplets which obscured visualization. This transport was greatly enhanced in the convective regime, and droplets visible to the eye would outline the underlying convection pattern on the window. A similar effect was found to take place when $\Delta T < 0$: the species contained within the upper layer would travel downwards and condense on the mirror. An image of the convection cell filled with Galden HT-70 and 5 cS silicone oil is shown in Fig. 16 at the end of an experimental run.

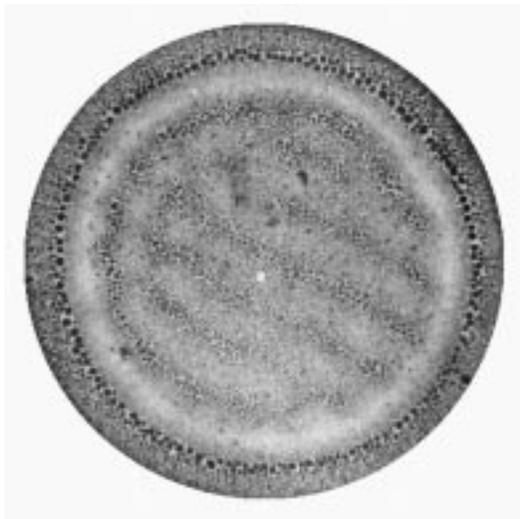

Fig. 16. Isothermal convection cell filled with Galden HT-70 and 5 cS silicone oil at the end of the experimental run with heating from above of Fig. 12(a). The shadowgraph is focused on the sapphire window. Large droplets of silicone oil are condensed on the mirror and weakly outline the pattern observed when heating from above. In addition, droplets of Galden HT-70, which were deposited when the sytem was heated from below, are wetting the sapphire window.

This droplet formation phenomenon was observed to various degrees in all the liquid combinations we tried, except in Galden HT-70 and water when the heating was applied from below. The Galden and silicone oil system comprises two polymerized liquids, while the acetonitrile and n-hexane are both composed of single molecules; the Galden and water combination is mixed. This suggests that the small water molecule can diffuse through the longer polymerized chains, but that the reverse process is not favored.

In order to reduce the droplet deposition onto the sapphire window in the acetonitrile and n-hexane system when heating from below, we attempted to alter the properties of the sapphire surface so that it would repel acetonitrile. The window was coated with silanes, which are strongly hydrophobic. In a second attempt, the surface of the sapphire was covered with a transparent teflon film of thickness 12.5 $\mu$m. Both procedures diminished the wetting of acetonitrile on the surface of the window. In addition, we observed a reduction in the deposition rate under experimental conditions, but the reduction was insufficient to allow for controlled experiments over periods of days.

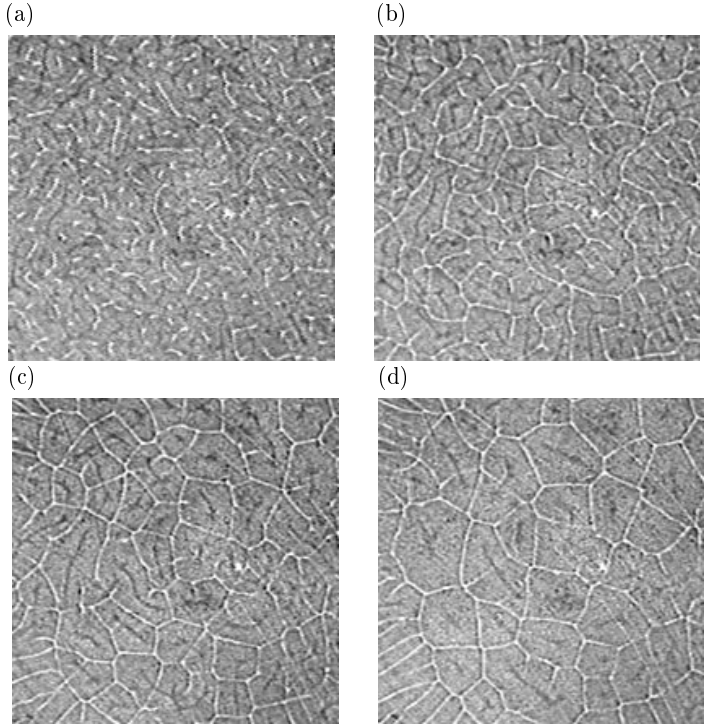

Fig. 17. Thermo-solutal-driven patterns observed when the direction of the applied temperature difference is reversed, after monitoring convection patterns for $\Delta T > 0$. Each image corresponds to the central 4.35 cm $\times$ 4.35 cm of the convection cell. The fluids are Galden HT-135 and 10 cS silicone oil ($d_{\text{tot}} = 4.25$ mm, $d_1 = 0.27\, d_{\text{tot}}$) and $\Delta T = -5.913°$C: (a) time = 5 mins, (b) time = 20 mins, (c) time = 45 mins, (d) time = 105 mins. The patterns are transient and their characteristic wavelength increases with time.

Generally, we were not able to reverse the deposition process and the experimental cell had to be dismantled and cleaned after short experimental runs. However, in the perfluorinated hydrocarbon and silicone oil experiments, the Galden liquid that attached to the sapphire window for $\Delta T > 0$ could be dislodged by reversing the direction of the applied temperature difference. In this case, a pattern of cells whose characteristic wavelength increased with time were observed within minutes on the shadowgraph, as shown in Fig. 17. This pattern is very similar to that observed by La Porta and Surko (27) of thermosolutal convection for $\Delta T < 0$. The pattern was transient and lasted for a period of several hours, depending on the strength of the applied temperature difference and on the amount of Galden condensed on the sapphire window.

Thus, the droplets of Galden were effectively mixed back into solution and travelled down into the bottom layer due to thermo-solutal forces.

The mixing effect was particularly surprising since water is soluble in Galden HT-70, for instance, only to 14 ppm. However, an estimate of the time scale required for a 0.5 $\mu$m thick water film to attach to the mirror, based on convective velocities at onset predicted by Engel and Swift (20), yielded 11 hours. This estimate provides an upper time limit for the suspected convection-enhanced diffusive process to become visible. However, properties such as wetting and temperature dependence of the solubility of the liquids need to be taken into account in order to reach a detailed understanding of this phenomenon.

## 5 Conclusion

Our study of Bénard-Marangoni convection in two liquid layers reveals a variety of convective phenomena. We have presented an overview of convection patterns observed for a selection of pairs of liquids and depths of layers. In particular, we have reported the first observation of Marangoni instability with the top plate hotter than the bottom plate and of oscillatory convection at onset when both thermocapillarity and buoyancy are significant.

## Acknowledgements

Authors WDM, JBS and HLS acknowledge many helpful discussions with John David Crawford, and remember especially his stay at the University of Texas in 1989, when he taught an excellent course entitled *Introduction to Bifurcation Theory*; the lecture notes for this course were published in (28).

The authors thank Steve Van Hook, Andreas Engel, Wayne Tokaruk and Stephen Morris for many useful discussions. This research is supported by the NASA Microgravity Science and Applications Division (Grant No. NAG3-1839) and the State of Texas Advanced Technology Program Grant 221 (G. Carey, P.I.).